\documentclass[a4j,12pt,onecolumn,oneside,notitlepage,final]{revtex4}
\usepackage{latexsym}
\usepackage{bm}
\usepackage{amsmath,amssymb}
\usepackage{mathrsfs}
\usepackage{tabularx}
\usepackage{float}
\usepackage{enumerate}
\usepackage[dvips]{graphicx}

\newcommand{\A}{\ddot{\rm{a}}}
\newcommand{\Pl}{\rm{Pl}}

\begin{document}
\begin{flushright}
   TU-795 
\end{flushright}
\thispagestyle{empty}
\begin{center}
{\Large{\bf{A Note on Polonyi Problem}}} \\ 
\end{center}
\begin{center}
Shuntaro Nakamura and Masahiro Yamaguchi \\
\textit{Department of Physics, Tohoku University, Sendai, 980-8578, Japan}
\end{center}
\begin{center}
{\large{Abstract}} 
\end{center} 
We reinvestigate the cosmological Polonyi problem in the case where the Polonyi mass is $\mathcal{O}(10) \, \rm{TeV}$. Such a large supersymmetry breaking scale implies that the Polonyi field should be sequestered from the standard model sector. Since the Polonyi field does not have a coupling to the
gauge multiplets at tree level, in order to obtain sufficiently high reheating temperature compatible with the standard big-bang nucleosynthesis the Polonyi mass well exceeds $100 - 1000$ TeV, depending on the 
decay channels.  
Moreover, we find that the branching ratio of the Polonyi field into neutralinos is of order unity, and thus the resulting neutralino LSPs, if stable, overclose the Universe even for the case of the wino-like LSP. Our explicit computation given here exhibits a very serious cosmological difficulty for models 
where supersymmetry breaking is caused by the Polonyi-type field.

\newpage
Supersymmetry (SUSY) is one of the most attractive candidates for new physics
beyond the standard model (SM), which solves the naturalness problem
associated with the electroweak scale.  Null results of superparticle
searches, however, require it to be broken. A natural thought
is that the supersymmetry is broken spontaneously in supergravity (SUGRA) 
\cite{Cremmer:1982en}, where
the super-Higgs mechanism is operative.  As a consequence, the would-be
Nambu-Goldstone fermion (goldstino) arises, which becomes the
longitudinal component of the gravitino. The properties of the scalar
superpartner of the goldstino are quite sensitive to how supersymmetry
is broken. In particular, when the supergravity effect plays an
essential role in supersymmetry breaking (termed the non-renormalizable
hidden sector in Ref. \cite{Banks:1993en}), the scalar superpartner of the 
goldstino often has interaction whose strength is similar to gravitational 
interaction, a mass, $m_\phi$, of the order of the gravitino mass, $m_{3/2}$, 
and a vacuum expectation value of the Planck scale. 
Such a field is called the Polonyi field \cite{Polonyi:1977}.

In this scenario, the lifetime of the Polonyi field, $\tau_\phi \sim
M_{\rm{Pl}}^2/m^3_\phi$, may become very long, where $M_{\rm{Pl}} \simeq
2.4 \times 10^{18} \, \rm{GeV}$ is the reduced Planck mass.  It is
well-known that the Polonyi field with the mass of the order of the
electroweak scale causes the cosmological difficulty.  In inflationary
epoch, the Polonyi field is shifted away from the true vacuum with a
magnitude, $\phi_{in}$, of the order of the Planck scale,
$\phi_{in} \simeq \mathcal{O}(M_{\rm{Pl}})$.  After inflation, the
Polonyi field starts a coherent oscillation around the true minimum when
the Hubble parameter becomes comparable to the Polonyi mass. Its energy
density will dominate the energy of the Universe soon after the start of
the oscillation. The subsequent decay of the Polonyi field releases
tremendous amount of entropy after nucleosynthesis, and thus it jeopardizes the
success of the big-bang nucleosynthesis (BBN).  This is the notorious Polonyi
problem \cite{Banks:1993en} \cite{Coughlan:1983ci} 
\cite{deCarlos:1993jw}.

It is known that the problem is somewhat ameliorated when the Polonyi
mass is as heavy as ${\cal O}(10)$ TeV
\cite{Moroi:1994rs} \cite{Kawasaki:1995cy} \cite{Moroi:1999zb}.  In this case, 
the reheating temperature by the Polonyi decay
would be sufficiently high so that its decay 
would not affect the BBN. The baryon number
asymmetry may be diluted at the reheating by the Polonyi decay, but can
still be as large as the observed value in the Affleck-Dine mechanism \cite{Affleck:1984fy}
(See, for instance, Ref. \cite{Kawasaki:2006py} \cite{Kawasaki:2007yy} for a recent work.). 
A difficulty in this case is the over-abundance of the lightest
superparticles (LSPs) which are produced at the Polonyi decay. The
annihilation among them does not work sufficiently, and thus too much
amount of the LSPs tends to remain in particular when the LSP is a
bino-like neutralino.

Recently, it was recognized that the gravitino production at the decay of the heavy scalar
field (the Polonyi field, the modulus field and the inflaton) 
is sizably large, which is incompatible with the big-bang
cosmology \cite{Nakamura:2006uc} \cite{Endo:2006zj} \cite{Asaka:2006bv}
\cite{Kawasaki:2006gs} \cite{Dine:2006ii} \cite{Endo:2006tf}.  
To avoid the gravitino overproduction in the
present situation, one should consider the case where the Polonyi decay
into the gravitino pair is kinematically forbidden, that is, $2 m_{3/2} > m_\phi$. 
Assuming that the gravitino mass is a few tens TeV or lighter, this condition implies that the Polonyi field cannot be
arbitrarily heavy, but can at most be as large as $100 \, \rm{TeV}$.  On
the other hand, with such a large supersymmetry breaking scale, the Polonyi field should be 
sequestered from the SM sector to keep the superparticle masses not very far from the electroweak scale. 
Absence of the direct coupling, however, raises the question how fast the Polonyi field can 
decay into the SM particles.

The purpose of this paper is to investigate the Polonyi decay and discuss its implications to
cosmology. The decay of heavy scalar fields, including an inflaton, via supergravity interaction 
was been investigated in detail in Refs. \cite{Nakamura:2006uc} \cite{Endo:2006zj}
\cite{Endo:2006qk} \cite{Endo:2007ih} \cite{Endo:2007sz}. 
Thus, we apply those results to this particular set-up.
Since the Polonyi field does not have a coupling to the
gauge multiplets at tree level, it will turn out that, 
 in order to obtain sufficiently high reheating temperature 
the Polonyi mass well exceeds 100 TeV, even if we consider efficient decay into 
right-handed neutrinos. 
In the absence of this decay channel, the Polonyi field dominantly decays into gauge multiplets. 
In this case, the Polonyi mass should be larger than 1000 TeV.  
We will also discuss the
partial decay into the superpartners in the SM sector. We find that
its branching ratio is generically of order unity, and thus the
resulting relic abundance of the neutralino LSPs, if stable, tends to be too
large, even when the LSP is a wino-like neutralino and its annihilation process 
is most efficient. 

To discuss the effect of the Polonyi field decay on the cosmology, we will consider the couplings of the Polonyi field to the various particles. If there were an unsuppressed coupling between the Polonyi field and the gauge kinetic term, the gaugino mass would be as large as $100 \, \rm{TeV}$. To avoid it, we assume that the gauge kinetic function does not depend on the Polonyi field $\phi$. 
Let us then discuss the coupling of $\phi$ to chiral multiplets. 
To obtain TeV scale soft scalar masses, we assume that the visible sector is sequestered from the Polonyi field, which implies the following form of the SUGRA $f$ function and the superpotential $W$
\begin{eqnarray} \label{eq: SUGRA f function}
   f &=& -3 \, \xi(\phi, \phi^*) + Q^{\dagger}_i e^{g V} Q_i, \\ 
   W &=& W(\phi) + \frac{1}{2} M N N + \frac{1}{6} Y_{ijk} Q^i Q^j Q^k,
\end{eqnarray}
where $\xi$ is an arbitrary function of $\phi$ and $\phi^*$, and $Q$'s denote matter fields. Here, $g$ and  $Y_{ijk}$ are the gauge coupling constant and the Yukawa coupling constant, respectively. 
We have introduced the right-handed neutrino, $N$, with mass $M$, which will play an essential role in our discussion.  
The function $f$ is related to the K$\A$hler potential, $K$, via
\begin{eqnarray} \label{eq: relation between f and K}
   f = -3 e^{-K/3},
\end{eqnarray}
where we used the Planck unit $M_{\rm{Pl}} = 1$. In what follows, we will use this unit unless we explicitly mention. 
From eq.$\,$\eqref{eq: relation between f and K}, we can obtain the K$\A$hler potential as
\begin{eqnarray} \label{eq: Kahler potential of phi}
   K = -3 \ln \xi(\phi, \phi^*) + \frac{1}{\xi} \, Q^{\dagger}_i e^{g V} Q_i + \cdots.
\end{eqnarray}
Now, we would like to discuss the decay of the Polonyi field. First, we consider the decay into the matter fermions. Relevant terms for producing the matter fermions are 
\begin{eqnarray} \label{eq:Lagrangian for fermions}
          \mathcal{L} &=& - i g_{i j^*} \bar{\chi}^j \bar{\sigma}^\mu D_\mu \chi^i 
                          + \frac{i}{4} g_{i j^*} 
                               \left( K_k \partial_\mu \phi^k - K_{k^*} \partial_\mu {\phi^*}^k \right)
                               \bar{\chi}^j \bar{\sigma}^\mu \chi^i 
                          - i g_{i j^*} \Gamma^i_{k \ell} 
                               \left( \partial_\mu \phi^k \right) 
                               \bar{\chi}^j \bar{\sigma}^\mu \chi^{\ell} \nonumber \\
                        &&- \frac{1}{2} e^{K/2} 
                               \left( \mathscr{D}_i \mathcal{D}_j W \right) \chi^i \chi^j + \rm{h.c.},
\end{eqnarray}
where  $\mathscr{D}_i \mathcal{D}_j W \equiv W_{ij} + K_{ij} W + K_i \mathcal{D}_j W + K_j \mathcal{D}_i W - K_i K_j W - \Gamma^k_{i j} \mathcal{D}_k W$ and $\mathcal{D}_i W \equiv W_i + K_i W$ with $\Gamma^k_{i j} \equiv g^{k \ell^*} g_{i j \ell^*}$. 
The K$\A$hler metric is $g_{i j^*} \equiv \partial^2 K/ \partial \phi^i \partial {\phi^*}^j$ and the covariant derivative $D_\mu = \partial_\mu - i g A_\mu^{(a)} T^{(a)}$ with the representation matrix, $T^{(a)}$, for the generator of the gauge group. The sum over indices is understood and $\phi^i$ represent all scalar fields including the Polonyi field. 
When the Polonyi field $\phi$ is written explicitly, $\phi^i$ and $\chi^i$ represent only the matter fields.     
From eq.$\,$\eqref{eq:Lagrangian for fermions} and using the equation of motion, we can obtain interactions of $\phi$ with chiral fermions \cite{Endo:2007sz}
\begin{eqnarray} \label{eq: interaction of phi with chi chi}
   \mathcal{L} = - \frac{1}{2} e^{K/2} 
                    \left( K_\phi W_{ij} - 2 \Gamma^k_{\phi i} W_{j k} \right)
                    \phi \chi^i \chi^j + \rm{h.c.},
\end{eqnarray}
where we set $K_i$, $W_i \ll 1$ because of the assumption that the matter fermions are charged under some symmetries. Eq.$\,$\eqref{eq: interaction of phi with chi chi} shows that the decay amplitude is proportional to the mass of a final-state fermion. Thus the decay into quarks and leptons in the SM sector are suppressed. By the same reason the three-body decay such as $\phi \to q \bar{q} g$ with 
$q$ and $g$ being a quark and a gauge boson, respectively, are also suppressed. 
On the other hand, the decay into the right-handed neutrino pair can be sizable, if it is kinematically allowed. The decay width  is 
\begin{eqnarray} \label{eq: decay width into neutrinos}
   \Gamma( \phi \to \nu_R \nu_R ) = 
      \frac{\lambda^2}{32 \pi} N_f \sqrt{1 - \frac{4 M^2}{m^2_\phi}}
      \left( 1 - \frac{2 M^2}{m^2_\phi} \right) \frac{M^2}{m^2_\phi} 
      \frac{m^3_\phi}{M^2_{\rm{Pl}}}, 
\end{eqnarray}
where we have written the Planck mass explicitly, and $\lambda \equiv \xi_\phi/\xi$ and $N_f$ is the number of the right-handed neutrinos. 

On the other hand, interactions of $\phi$ with matter scalars come from the kinetic term and the scalar potential  
\begin{eqnarray} \label{eq: Lagrangian for scalars}
   \mathcal{L} = - g_{i j^*} D_\mu \phi^i D^\mu {\phi^*}^j 
                 - e^K \left[ g^{i j^*} ( \mathcal{D}_i W ) ( \mathcal{D}_j W )^* - 3 |W|^2 \right].
\end{eqnarray}
We can obtain the interactions for the decay $\phi \to \phi^i {\phi^*}^j$ and $\phi \to \phi^i {\phi^*}^j g$  from the kinetic term of eq.$\,$\eqref{eq: Lagrangian for scalars}. As in the case of fermions, one can also check that these decay widths are suppressed by the masses of the final-state scalars. Thus, they are also subdominant component of the total decay width of $\phi$ unless the final-state scalars have quite large soft SUSY breaking masses. 
The interactions corresponding to $\phi \to \phi^i \phi^j$ are obtained as
\begin{eqnarray}
   \mathcal{L} = - \frac{1}{2} e^K \left( K_\phi W_{ij} - 2 \Gamma^k_{\phi i} W_{jk} \right)^*
                   g^{\phi \phi^*} W_{\phi \phi} \, \phi \, {\phi^*}^i {\phi^*}^j + \rm{h.c.}, 
\end{eqnarray}
where $g^{\phi \phi^*} e^{K/2} W_{\phi \phi} \equiv m_\phi$. Therefore, the decay into the right-handed sneutrino pair is the dominant decay  channel for producing a pair of matter scalars, if it is allowed kinematically. The decay width is calculated as
\begin{eqnarray} \label{eq: decay width into sneutrinos}
   \Gamma( \phi \to \tilde{\nu}_R \tilde{\nu}_R ) = 
      \frac{\lambda^2}{128 \pi} N_f \sqrt{1 - \frac{4 M^2}{m^2_\phi}} 
      \frac{M^2}{m^2_\phi} \frac{m^3_\phi}{M^2_{\rm{Pl}}}.
\end{eqnarray}

The Polonyi field may decay into three-body final states such as $\phi \to \phi^i \chi^j \chi^k$ and $\phi \to \phi^i \phi^j \phi^k$. The former process occurs through the interactions as \cite{Endo:2007sz}
\begin{eqnarray} \label{eq: Lagrangian for 1scalar and 2fermions}
   \mathcal{L} = - \frac{1}{2} e^{K/2}
                   \left( K_\phi W_{ijk} - 3 \Gamma^{\ell}_{\phi i} W_{jk \ell} \right)
                   \phi \, \phi^i \chi^j \chi^k + \rm{h.c.},
\end{eqnarray}
and the three-scalar final state process is obtained by 
\begin{eqnarray} \label{eq: Lagrangian for 3scalars}
   \mathcal{L} = - \frac{1}{6} e^K 
                   \left( K_\phi W_{ijk} - 3 \Gamma^{\ell}_{\phi i} W_{jk \ell} \right)^*
                   g^{\phi \phi^*} W_{\phi \phi} \phi \, {\phi^*}^i {\phi^*}^j {\phi^*}^k 
                 + \rm{h.c.}.
\end{eqnarray}
However, from the K$\A$hler potential \eqref{eq: Kahler potential of phi}, the parentheses in eq.$\,$\eqref{eq: Lagrangian for 1scalar and 2fermions} and \eqref{eq: Lagrangian for 3scalars} vanish. The three-body decay process, then, does not occur. 

In addition to these tree-level decay processes, the Polonyi field can also decay into the gauge supermultiplets through the anomaly-mediation effect \cite{Endo:2007ih}, even if $\phi$ does not have any direct couplings to the gauge sector. Taking account of only the strong coupling, we obtain  \cite{Endo:2007ih}
\begin{eqnarray} \label{eq: decay width of phi with anomaly mediation}
    \Gamma^{\rm{AM}}_{(\phi \to g g)} = \Gamma^{\rm{AM}}_{(\phi \to \tilde{g} \tilde{g})}
       \simeq \frac{9 \alpha^2_s}{128 \pi^3} \lambda^2 
              \frac{m^3_\phi}{M^2_{\rm{Pl}}}.
\end{eqnarray}

If the decay processes $\phi \to \nu_R \nu_R$ and $\phi \to \tilde{\nu}_R \tilde{\nu}_R$ are 
not allowed kinematically, the dominant contribution to reheat the Universe comes from 
the anomaly-induced decay eq.$\,$\eqref{eq: decay width of phi with anomaly mediation}. 
However, it is unlikely that such a decay width provides sufficiently high reheating temperature, 
since it is suppressed by a loop factor.
On the other hand, if $\phi$ can decay into the right-handed (s)neutrino pair, there is a possibility 
that the process of decay into the right-handed (s)neutrino pair is more efficient than the 
anomaly-induced one. The possibility is that we will tune the right-handed (s)neutrino mass $M$ to 
maximize the decay width. 
In that case, the total decay width of $\phi$ comes from eq.$\,$\eqref{eq: decay width into neutrinos} and eq.$\,$\eqref{eq: decay width into sneutrinos}, that is, 
\begin{eqnarray} \label{eq: total decay width}
   \Gamma_{\rm{tot}.} &\simeq& 
      \Gamma(\phi \to \nu_R \nu_R) + \Gamma(\phi \to \tilde{\nu}_R \tilde{\nu}_R) \nonumber \\ 
         && = \frac{N_f}{32 \pi} \lambda^2 \sqrt{1 - \frac{4 M^2}{m^2_\phi}} 
              \left( \frac{5}{4} - \frac{2 M^2}{m^2_\phi} \right) \frac{M^2}{m^2_\phi}
              \frac{m^3_\phi}{M^2_{\rm{Pl}}}.
\end{eqnarray}
When the right-handed (s)neutrino mass $M$ satisfies the relation $M \simeq 0.38 \, m_\phi$, the total 
decay width is maximized as 
\begin{eqnarray} \label{eq: maximized total decay width}
   \Gamma^{\rm{max}}_{\rm{tot}.} \simeq 8.9 \times 10^{-3} N_f \lambda^2 \frac{m^3_\phi}{M^2_{\rm{Pl}}}.
\end{eqnarray}
From eq.$\,$\eqref{eq: maximized total decay width}, the reheating temperature, $T_R(\phi)$, after $\phi$ decay is 
\begin{eqnarray}  \label{eq: reheating temp.}
   T_R(\phi) &\equiv& \left( \frac{90}{\pi^2 g_*} \right)^{1/4} 
                   \sqrt{ \Gamma^{\rm{max}}_{\rm{tot.}} M_{\Pl} } \nonumber \\ 
               &=& 1.9 \times 10^{-3} \, {\rm{GeV}} \, \lambda \sqrt{N_f}
                   \left( \frac{g_*}{10} \right)^{-1/4} 
                   \left( \frac{m_\phi}{10^5 \, \rm{GeV}} \right)^{3/2}, 
\end{eqnarray}
where $g_*$ is the effective degrees of freedom of the radiation at the reheating. 
In Ref. \cite{Kawasaki:2000en}, it was shown that in order to reproduce the observed abundance of $^4$He, the reheating temperature should be higher than about $4-7$ $\rm{MeV}$ \footnote{The definition of the reheating temperature in Ref. \cite{Kawasaki:2000en} is different from our definition by the factor  $\sqrt{3}$.} for the hadronic branching ratio $B_h = 10^{-2} - 1$. We find that $m_\phi$ should be heavier than $3 \times 10^5$ GeV for $\lambda = N_f =1$ so that the reheating temperature 
survives this bound.
Notice that in the absence of decay channels into right-handed (s)neutrino pair, 
the decay is dominated by eq.$\,$\eqref{eq: decay width of phi with anomaly mediation}. 
In this case, the Polonyi mass should be heavier than about $2 \times 10^6$ GeV. 
This implies that the gravitino mass is heavier than $10^6$ GeV, which is out of the region of the 
low-energy supersymmetry. 
 
\begin{figure}[h]
   \includegraphics[width=9cm, clip]{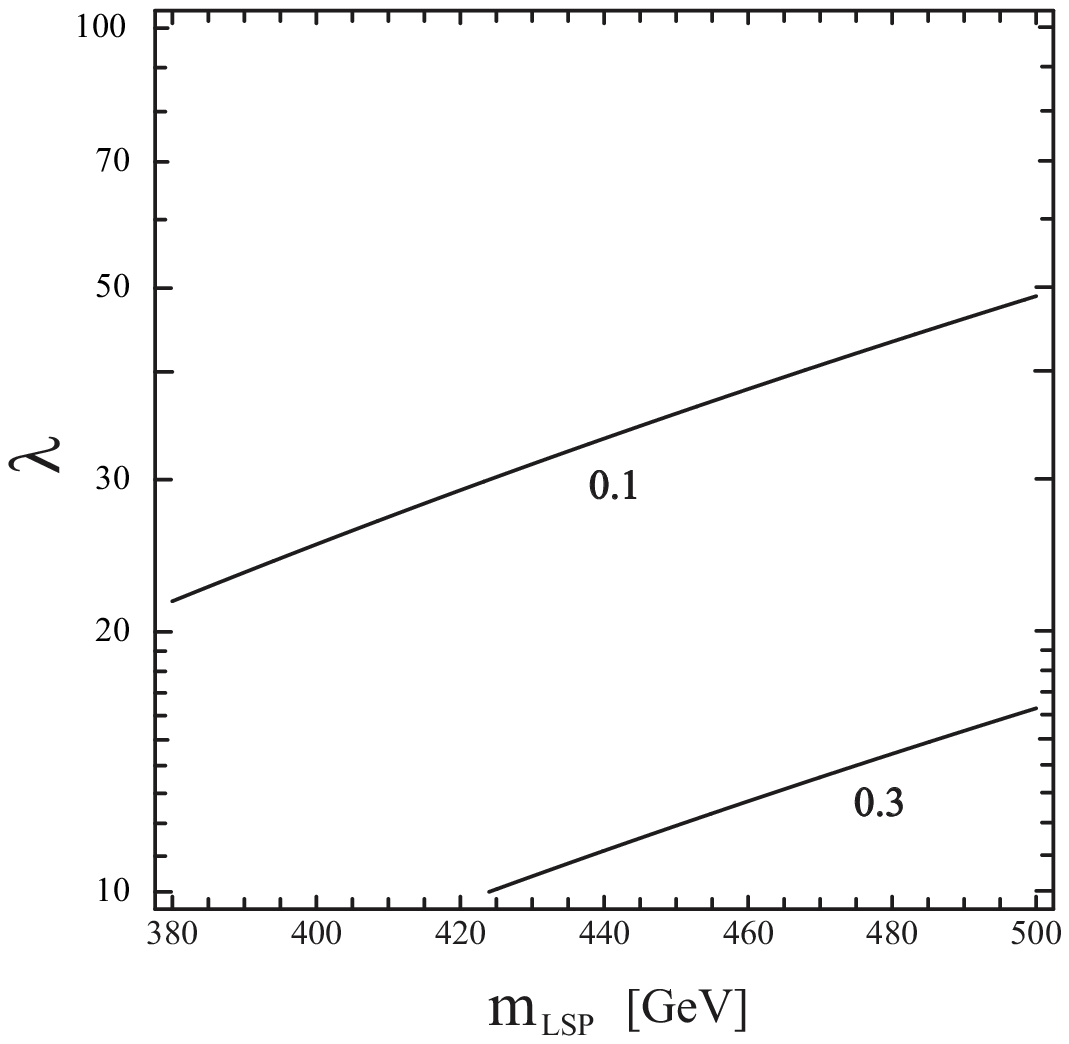} 
   \caption{Contours of the density parameter $\Omega_{\rm{LSP}} h^2$ for the wino-like LSP drawn in $m_{\rm{LSP}}- \lambda$ plane. Two lines represent $\Omega_{\rm{LSP}} h^2 = 0.1, ~ 0.3$, from the above.} 
\label{fig: contour of LSP abundance}
\end{figure}

Let us next consider a more stringent constraint imposed by the neutralino LSP relic abundance. 
Neutralino LSPs are produced by subsequent decay of the right-handed sneutrino, as well as  through the anomaly-mediated decay eq.$\,$\eqref{eq: decay width of phi with anomaly mediation}. Since the branching ratio of $\phi$ into neutralino LSPs is $\mathcal{O}(1)$, they are so abundant that the annihilation among them becomes effective. The annihilation process will cease when the Hubble parameter becomes comparable to the annihilation rate
\begin{eqnarray}
   \langle \sigma_{\rm{ann.}} v_{\rm{rel.}} \rangle n_{\rm{LSP}} \simeq H(T_R),
\end{eqnarray}
where $\sigma_{\rm{ann.}}$ is the annihilation cross section of two LSPs, $v_{\rm{rel.}}$ their relative velocity, $\langle \cdots \rangle$ represents the thermal average, and $n_{\rm{LSP}}$ is the number density of the LSPs. Thus, the yield of the LSPs at the $\phi$ decay can be estimated as 
\begin{eqnarray} \label{eq: yield of LSP}
   Y_{\rm{LSP}}^{\rm{ann.}} &\simeq& 
      \frac{H(T_R)}{\langle \sigma_{\rm{ann.}} v_{\rm{rel.}} \rangle s}
      = \frac{1}{4} \left( \frac{90}{\pi^2 g_*(T_R)} \right)^{1/2}
        \frac{1}{\langle \sigma_{\rm{ann.}} v_{\rm{rel.}} \rangle T_R M_{\Pl}}. 
\end{eqnarray}
When we consider the case where the wino is the LSP, the annihilation process is most effective. The annihilation cross section is obtained as \cite{Olive:1989jg}
\begin{eqnarray} \label{eq: annihilation cross section of wino}
   \langle \sigma_{\rm{ann.}} v_{\rm{rel.}} \rangle 
      = \frac{g^4_2}{2 \pi} \frac{1}{m^2_{\rm{LSP}}} \frac{(1-x_W)^{3/2}}{(2-x_W)^2}, 
\end{eqnarray}
where $g_2$ is the $SU(2)$ gauge coupling constant, $m_{\rm{LSP}}$ the LSP mass, and $x_W \equiv m^2_W/m^2_{\rm{LSP}}$ with $W$ boson mass $m_W$. From eq.$\,$\eqref{eq: yield of LSP} and eq.$\,$\eqref{eq: annihilation cross section of wino}, we can compute the ratio of the LSP mass density to the entropy density: 
\begin{eqnarray} \label{eq: LSP energy density}
  m_{\rm{LSP}} \frac{ n_{\rm{LSP}} }{s} 
      &\simeq& 1.9 \times 10^{-9} \, {\rm{GeV}} \frac{1}{\lambda \sqrt{N_f}} 
               \, \frac{(2-x_W)^2}{(1-x_W)^{3/2}} 
               \left( \frac{m_{\rm{LSP}}}{380 \, \rm{GeV}} \right)^3 \nonumber \\
      &&\times \left( \frac{g_*}{10} \right)^{-1/4} 
               \left( \frac{m_\phi}{3 \times 10^5 \, \rm{GeV}} \right)^{-3/2},
\end{eqnarray}
where we have set the wino mass equal to 380 GeV. The reason is that since the Polonyi decay into the gravitino pair must be forbidden kinematically, gravitinos have to be heavier than $1.5 \times 10^5$ GeV. Such a heavy gravitino induces the wino mass about $m_{\widetilde{W}} \simeq 380$ GeV via the anomaly mediated SUSY breaking effect \cite{Randall:1998uk}.
It is convenient to write eq.$\,$\eqref{eq: LSP energy density} in terms of the $\Omega$ parameter which is defined by the ratio of the LSP mass density to the critical mass density, 
\begin{eqnarray} \label{eq: Omega of LSP}
   \Omega^{\rm{ann.}}_{\rm{LSP}} h^2 \simeq 
             0.53 \, \frac{1}{\lambda \sqrt{N_f}} \frac{(2-x_W)^2}{(1-x_W)^{3/2}} 
             \left( \frac{m_{\rm{LSP}}}{380 \, \rm{GeV}} \right)^3 
             \left( \frac{g_*}{10} \right)^{-1/4} 
             \left( \frac{m_\phi}{3 \times 10^5 \, \rm{GeV}} \right)^{-3/2},
\end{eqnarray}
where $h \simeq 0.72$ is the Hubble constant in units of 100 km/Mpc/s. The contours of eq.$\,$\eqref{eq: Omega of LSP} in $m_{\rm{LSP}} - \lambda$ plane are shown in fig.$\,$\ref{fig: contour of LSP abundance}. 
Recent WMAP observations \cite{Spergel:2006hy} suggest that the density parameter of the dark matter be $\Omega_{\rm{DM}} h^2 = 0.105^{+ 0.007}_{-0.013}$ (68 \% C.L.). 
Therefore, from fig.$\,$\ref{fig: contour of LSP abundance}, in order to avoid the LSP overclosure, $\lambda \gtrsim 20$ even for the wino-like LSP.   
Since $\lambda$ is the coupling constant between the Polonyi field and the matter fields, it is natural to expect that it is of order unity. Thus, even if the LSP is the wino, it cannot explain the present dark matter abundance. This conclusion also applies to the case where the LSP is bino- and higgsino-like one because the annihilation cross section is even smaller.

In summary, we have reconsidered the cosmological implications of the heavy Polonyi field with the mass $\mathcal{O}(10) \, \rm{TeV}$. To avoid the heavy gaugino mass, the gauge kinetic function is assumed to be independent of the Polonyi field. In such a case, even when the Polonyi field can decay into the right-handed (s)neutrino pair and we have tuned the right-handed (s)neutrino mass $M$ to maximize the 
total decay width, we found that the Polonyi mass well exceeds 100 TeV in order to obtain sufficiently high reheating temperature compatible with the standard BBN. 
In the absence of this decay channel, the Polonyi field dominantly decays into gauge multiplets through the anomaly mediation effect. 
In this case, however, the Polonyi mass has to be heavier than about 1000 TeV. 
We also discussed the neutralino LSP abundance produced by the Polonyi decay. We found that, if the neutralino LSP is stable,  avoidance of the LSP overclosure requires $\lambda \gtrsim 20$ even when the LSP is a wino-like neutralino and its annihilation process is most efficient. 
This result implies that even for the wino-like LSPs, its abundance produced by the Polonyi decay cannot account for the present dark matter abundance. Other types of the neutralino LSPs would also be too abundant to be consistent with the WMAP observations.  
All in all, the explicit computation presented here makes the Polonyi problem even worse, 
and thus one should probably consider supersymmetry breaking scenarios in the absence of the 
Polonyi-type field.

\section*{{ACKNOWLEDGMENTS}}
We would like to thank T. Higaki for useful discussions. The work was partially supported by the grants-in-aid from the Ministry of Education, Science, Sports, and Culture of Japan, No.16081202 and No.17340062.

\newpage

\end{document}